# Out-of-equilibrium electrons lead to record thermionic emission in LaB$_6$ with the Jahn-Teller instability of boron cage

*Elena S. Zhukova[1,2,\*], Boris P. Gorshunov[1,2,\*], Martin Dressel[1,3], Gennadii A. Komandin[2], Mikhail A. Belyanchikov[1], Zakhar V. Bedran[1], Andrei V. Muratov[4], Yuri A. Aleshchenko[4], Mikhail A. Anisimov[2], Nataliya Yu. Shitsevalova[5], Anatoliy V. Dukhnenko[5], Volodymyr B. Filipov[5], Vladimir V. Voronov[2] and Nikolay E. Sluchanko[1,2,\*]*

[1-]Moscow Institute of Physics and Technology, 9 Institutskiy Pereulok, 141700 Dolgoprudny, Moscow Region, Russia

[2-]Prokhorov General Physics Institute of the Russian Academy of Sciences, 38 Vavilov Street, 119991 Moscow, Russia

[3-]Physikalisches Institut, Universita¨t Stuttgart, 57 Pfaffenwaldring, D-70569 Stuttgart, Germany

[4-]Lebedev Physical Institute, Russian Academy of Sciences, 53 Leninskiy Avenue, 119991 Moscow, Russia

[5-] Frantsevich Institute for Problems of Materials Science, National Academy of Sciences of Ukraine, 3 Krzhyzhanovsky Street, 03680 Kiev, Ukraine

[\*]Email: Zhukova.es@mipt.ru, gorshunov.bp@mipt.ru, nes@lt.gpi.ru

**Abstract**

Materials with low electron work function are of great demand in various branches of science and technology. LaB$_6$ is among the most effective electron-beam sources with one of the highest brightness of thermionic emission. A deep understanding of the physical mechanisms responsible for the extraordinary properties of LaB$_6$ is required in order to optimize the parameters and design of thermionic elements for application in various electron-beam devices. Motivated by recent experiments on rare earth borides indicating a strong coupling of conduction electrons to the crystal lattice and rare earth ions, we have studied the state of electrons in the conduction band of lanthanum hexaboride by performing infrared spectroscopic, DC resistivity and Hall-effect studies of LaB$_6$ single crystals with different ratios of $^{10}$B and $^{11}$B isotopes. We find that only a small amount of electrons in the conduction band behave as Drude-type mobile charge carriers while up to 70% of the electrons are far out of equilibrium and involved in collective oscillations of electron density coupled to vibrations of the Jahn-Teller unstable rigid boron cage and rattling modes of La-ions that are loosely bound to the lattice. We argue that exactly these non-equilibrium (hot) electrons in the conduction band determine the extraordinary low work function of thermoemission in $LaB_6$. Our observation may guide future search for compounds with possibly lower electron work function.

1. **Introduction**

Materials with a low work function find diverse applications in electronic devices where they are utilized as contact electrodes or electronic emitters (see, e.g.[1]). Due to the high electronic emissivity, melting point, mechanical and chemical stability, rare earth (RE) hexaborides $RB_6$ ($R$ is a metal ion) are among the most promising compounds for high-power electronic technology. There still is a lack of theoretical understanding of the microscopic mechanisms that determine these remarkable characteristics. RE hexaborides reveal a rich variety of interesting physical properties. Among them are metals ($LaB_6$-$NdB_6$, $GdB_6$-$HoB_6$)[2-4], semimetal ($EuB_6$)[5], semiconductor ($YbB_6$)[6], superconductor ($YB_6$)[7] and intermediate valence Kondo insulator ($SmB_6$)[8]. Such diversity is determined by specific features of crystal structure and electronic configuration of the RE-ions. These particular electronic and ionic properties of the RE hexaborides can be assumed to be at the origin of their unique thermionic emission ability. Their crystal structure is shown in the inset in Figure 1a. It has the bcc $CaB_6$-type with Pm3m-$O_1^h$ symmetry and contains two types of atoms, RE and boron, where the boron ions are organized into $B_6$ clusters (octahedrons). The structure can be represented as a rigid network of covalent-bounded octahedral-shaped $B_6$ complexes with RE ions embedded in cavities formed by $B_6$ clusters[9]. The metal atoms are located within the $B_{24}$ polyhedra and are loosely bound to the surrounding boron atoms, both factors leading to quasi-local vibrations (rattling modes)[10]. It was found recently that in the higher borides $R$B$_{12}$, development of the Jahn-Teller (JT) instability in the boron B$_{12}$ clusters leads to emergence of an infrared-active collective excitation that involves corresponding JT-mode, produces the rattling vibration of the RE atom and modulation of electronic density in the conduction band[11-12]. Very similar collective excitations were discovered in Gd$_x$La$_{1-x}$B$_6$ as well[13]. It was suggested that they arise from the complex interaction among lattice, orbital and charge carriers subsystems, and that their striking consequence is the conversion, by virtue of collective interactions, of large fraction of conduction electrons into a non-equilibrium state with strong scattering. This phenomenon was also proposed[13] to cause a record low thermal emission work function of LaB$_6$ ($\varphi \approx 2.66$ eV[14]).

Since the boron network is essentially involved in the formation of the collective excitations – via the cooperative-dynamic JT-effect on the B$_6$ clusters –, here our goal was to explore in detail its evolution and in this way the evolution of the conduction electrons state upon isotopic substitution in the boron sublattice. We use infrared spectroscopy (frequency range 40 - 35000 cm$^{-1}$) together with DC resistivity and Hall effect measurements to study the electronic properties of several isotopically substituted La($^{10}$B$_x$$^{11}$B$_{1-x}$)$_6$ compounds with x=0, 0.189, 0.5, 0.75, 1. In addition to the Drude-type free carrier component, we find that the infrared spectra of all studied compounds exhibits distinct signatures of a collective excitation with complex, slightly non-Lorentzian lineshape and unusually large dielectric contributions reaching values $\Delta\varepsilon$=7300 for La$^{10}$B$_6$ and $\Delta\varepsilon$=4050 for La$^{11}$B$_6$. Our analysis reveals that up to about 70% of conduction band electrons are involved in the formation of the excitation; with other words, they exist in some kind of non-equilibrium state. The largest fraction of such electrons is detected in lanthanum hexaboride with natural isotopic composition (La$^{nat}$B$_6$) where the excitation has the highest damping due to disorder within the boron sub-lattice. We argue that such disorder is an important factor driving the system of conduction electrons of LaB$_6$ to the non-equilibrium state and that these "hot" electrons determine the exceptionally low work function of the compound.



## 2. Experimental details

High quality La($^{10}$B$_x^{11}$B$_{1-x}$)$_6$ single crystals with $x$=0, 0.189 (La$^{nat}$B$_6$), 0.5, 0.75 and 1 were grown by vertical crucible-free inductive zone melting in argon gas atmosphere. The details of crystal preparation are described in[15]. With the exception of the isotopically pure $^{11}$B, the intended isotope enrichment was adjusted by the respective mixtures of $^{10}$B and $^{nat}$B, accordingly. The samples with the lowest concentrations of isotopic impurities are La($^{10}$B$_{0.971}^{11}$B$_{0.029}$)$_6$ and La($^{10}$B$_{0.005}^{11}$B$_{0.995}$)$_6$, taking into account the actual enrichment of the used individual isotopes; for convenience these two samples are designated as the isotopically pure materials. The grown crystals were characterized by recording Laue back-patterns, optical spectral analysis, magnetization, DC transport and Hall effect measurements. Apart from the boron 'isotopic impurities', the total impurity content of our samples is less than $10^{-3}$ mass %. For the infrared reflectivity measurements the surfaces of 5*5 mm$^2$ area samples were made plane within ±1 μm and polished with diamond powder. Finally all samples were etched in dilute nitric acid (HNO$_3$:H$_2$O = 1:2, 10 s) to avoid distortions of the surface layer due to polishing.

In the frequency range $v$=40-8000 cm$^{-1}$, the reflectivity spectra $R(v)$ were measured employing a Bruker Vertex 80V Fourier-transform infrared (IR) spectrometer; gold films deposited on a glass substrate were used as reference mirrors. With the J.A. Woollam V-VASE ellipsometer, spectra of optical conductivity $\sigma(v)$ and dielectric permittivity $\varepsilon'(v)$ of the samples were directly determined in the interval 3700 cm$^{-1}$ – 35000 cm$^{-1}$. From the ellipsometry data, the reflection coefficients were calculated and merged with the measured IR reflectivity spectra. The data from Ref. [16] were taken to extend the spectral range up to ≈400000 cm$^{-1}$. DC conductivity $\sigma_{DC}$ and Hall resistivity of the same samples were measured using a standard 5-probe method.

## 3. Experimental results and discussion

In Figure 1a, the room-temperature reflectivity spectra of La$^{10}$B$_6$ and La$^{11}$B$_6$ crystals are shown in a large the frequency range The overall spectra are typical for a good metal: a pronounced



plasma edge is observed at $\nu \approx 17000$ cm$^{-1}$ with a rather high reflectivity below. Features above 20000 cm$^{-1}$ are caused by electronic interband transitions[16-18] and will not be discussed here. The optical response of metals is commonly analyzed by the Drude model of the complex conductivity [19]

$$\sigma^*_{\text{Drude}}(\nu) = \frac{\sigma^{\text{Drude}}_{\text{DC}}}{1 - i\nu/\gamma^{\text{Drude}}}, \qquad (1)$$

where $\sigma^{\text{Drude}}_{\text{DC}}$ is the DC conductivity and $\gamma^{\text{Drude}}$ is the charge carriers scattering rate. For the example of La$^{11}$B$_6$, Fig. 1a demonstrates a fit (dashed line) of the data by Eq. (1) using $\gamma^{\text{Drude}}$=260 cm$^{-1}$ and the measured $\sigma^{\text{Drude}}_{\text{DC}}$=143 110 ($\Omega$ cm)$^{-1}$. Already above a few hundred cm$^{-1}$, significant deviations occur that are also identified in the spectra of the other crystals. The observed non-Drude shape of the reflectivity spectra cannot be explained by electronic correlations. A corresponding enhancement of the effective electronic mass is commonly confined to low temperatures, as observed in heavy fermions [20]. Concomitantly, applying the generalized Drude analysis to our data in order to determine the frequency dependence of the effective mass and scatting rate[19], we extract at the lowest frequencies (below 100-300 cm$^{-1}$) a carrier mass $m^*$ that falls in the range of $(0.9 \div 1)m_0$ for all studied compounds ($m_0$ is a free electron mass) indicating absence of strong correlations. We thus conclude that the infrared response of all studied La($^{10}$B$_x$$^{11}$B$_{1-x}$)$_6$ crystals is determined not just by free charge carriers – providing their relatively high metal-like conductivity – but that additional mechanisms contribute to the electronic properties of the compounds and produce non-Drude shape of the reflectivity spectra. From the intensity, width and shape rules of the features, simple lattice vibrations can be ruled out as possible origin of these bands. We formally model these mechanisms by introducing, in addition to the Drude free carrier term (1), a minimal set of excitations that provide description of the measured reflectivity spectra. We use Lorentzian expression

$$\sigma^*(\nu) = \frac{0.5 f \nu}{\nu \gamma + i(\nu_0^2 - \nu^2)} \qquad (2),$$

where $v_0$ is the resonance frequency, $f = \Delta\varepsilon v_0^2$ is the oscillator strength, $\Delta\varepsilon$ is the dielectric contribution and $\gamma$ is the damping constant. For all studied samples, it was sufficient to introduce two Lorentzians to perfectly fit the measured reflectivity spectra; examples of such fits obtained for La$^{10}$B$_6$ and La$^{11}$B$_6$ are presented by solid lines in Fig.1a. The so-obtained conductivity spectra for all the different samples La($^{10}$B$_x$$^{11}$B$_{1-x}$)$_6$, are shown in Fig.1b, where one can identify – next to the Drude roll-off – a bump-like feature around 500 cm$^{-1}$ that is present in all compositions x = 0, 0.189, 0.5, 0.75 and 1. Fig.2 displays separately the two contributions – from the Drude-type conduction electrons and from the peak feature. One can see that except for La$^{11}$B$_6$ all peaks are located at almost same frequency and that for the *x*=0 compound (La$^{11}$B$_6$) with *heavier* $^{11}$B ions the peak shifts towards *higher* frequencies. Below we account for this unusual behavior in terms of JT-distortions of boron clusters, which are the trigger for occurrence of collective mode in higher borides.[21]

Characteristics of the observed peaks - oscillator strengths *f*, dielectric contributions $\Delta\varepsilon$ and damping parameters $\gamma_{peak}$ vary moderately and non-monotonously with *x* in the La($^{10}$B$_x$$^{11}$B$_{1-x}$)$_6$ crystals reaching maximal values for La$^{nat}$B$_6$. In Figure 3 we summarize all parameters of the two contributions (from Lorentzians and Drude responses) obtained by fitting the reflectivity spectra. The discovered excitations have a rather unusually large dielectric contributions $\Delta\varepsilon$ ranging from $\Delta\varepsilon \approx 4000$ for La$^{11}$B$_6$ up to almost $\Delta\varepsilon \approx 8000$ for La$^{nat}$B$_6$, and are strongly overdamped (relative damping constants $\gamma/v_0 = 1 \div 3.4$). It is worth noting that the results we obtain here for the La($^{10}$B$_x$$^{11}$B$_{1-x}$)$_6$ are qualitatively similar to those deduced from the infrared experiments on Gd$_x$La$_{1-x}$B$_6$ hexaborides[13] and LuB$_{12}$,[11] Tm$_{0.189}$Yb$_{0.811}$B$_{12}$ dodecaborides.[12] In these compounds, in addition to the free-carrier Drude spectral component, strong infrared excitations were discovered with dielectric contributions $\Delta\varepsilon$ reaching values as high as 15000 and with non-Lorentzian lineshapes.

In Ref. [11-13] the origin of the excitations was associated with cooperative-dynamic Jahn-Teller effect in the boron sub-lattices, which produces quasi-local vibrations (rattling modes) of loosely bound RE ions, leading to 'modulation', via hybridization of 5*d*-conduction electrons and



$2p$-boron states, of the conduction band along certain crystallographic directions. We believe that similar mechanisms are responsible for the peak observe in the present study of La($^{10}$B$_x$$^{11}$B$_{1-x}$)$_6$ crystals. More specifically, because of double orbital degeneracy of the highest occupied molecular orbital, the B$_6$ molecules are JT active and their structure is thus labile due to JT distortions. In such cases, certain intrinsic structural defects (e.g. boron vacancies and/or mixed $^{10}$B/$^{11}$B isotope content in B$_6$ molecules) lift the degeneracy and lower the B$_6$ symmetry. Taking into account that B$_6$ clusters in hexaborides form an extended 3D rigid network connected by B-B covalent bonds, we suggest that the structural JT liability retains and is reinforced in the boron sub-lattice of $R$B$_6$. The reinforcement due to cooperative dynamic JT effect manifested both in static and dynamic lattice properties may be considered as the cause of large amplitude displacements of La atoms in oversized B$_{24}$ cages, resulting in both, distortions of the $bcc$ lattice and emergence of the rattling modes attributed to quantum motion of the La ions in the double-well potentials (DWPs) with the minima separated by ~0.5 Å (N.B.Bolotina, private communication). The corresponding barrier height of $\Delta E/k_B$~90 K was extracted from heat capacity measurements[22] of LaB$_6$ with different $^{10}$B/$^{11}$B isotope composition ($k_B$ is Boltzmann constant). The large amplitude vibrations of heavy ions in the double-well potentials are sketched in the inset in Figure 1b. As a consequence of the quantum motion (zero temperature vibrations) of La-ions, dramatic changes of the $5d - 2p$ hybridization of electron states should occur in hexaborides resulting in (*i*) the formation of a collective mode (overdamped oscillator) and (*ii*) the emergence of non-equilibrium (hot) charge carriers.

Taking for LaB$_6$ the value of the electrons effective mass $m^*$=0.6$m_0$[23] and using the relations for charge carriers plasma frequency $v_{pl}$=[$ne^2/(\pi m^*)$]$^{1/2}$ ($n$ is the concentration of electrons, $e$ – their charge) and oscillator strength of the Lorentzians $f = \Delta\varepsilon v_0^2 = ne^2(\pi m^*)^{-1}$, for every studied crystal we estimate the concentration of (*i*) free carriers that participate in the Drude conductivity ($n_{\text{Drude}}$), and (*ii*) electrons involved in the formation of the peak ($n_{\text{peak}}$). Furthermore, we calculate the mobility $\mu = \frac{e\tau}{m^*} = e(2\pi m^* \gamma^{\text{Drude}})^{-1}$ and the mean-free path $l=v_F\tau$ of the carriers responsible for the Drude charge transport (here $\tau$ is the relaxation time, $\gamma^{Drude}$=90 cm$^{-1}$ for all crystals studied, and we use the Fermi velocity $v_F \approx 6*10^7$ cm/s in LaB$_6$ found in[23]). The obtained data are collected in Figure 3. It is worth noting that the value of the total electronic concentration $n_{Drude} + n_{peak}$ found in the present optical experiments is in accordance with the concentration $n_{Hall}=(R_H e)^{-1}$ obtained from our Hall effect measurements on the same samples ($R_H$ is the Hall coefficient) thus confirming the validity of our calculations. Moreover, the obtained total concentration of electrons ~ (1.4-1.6)*10$^{22}$ cm$^{-3}$ coincides well with the number of La$^{3+}$ ions per unit volume in LaB$_6$ [$n$(La)~1.4*10$^{22}$ cm$^{-3}$, see dashed horizontal line in Figure 3b], in agreement with the well-known consideration of hexaborides as single electron metals (see e.g.[2]). It is interesting that the mobility determined from the DC Hall effect and resistivity measurements ($\mu_{Hall}$) is much lower than those found for the free (Drude) electrons, but, at the same time, much higher than the values found for electrons that participate in the formation of an overdamped peak ($\mu_{peak}$, see Figure 3c). This means that the Hall experiments determine the mobility of *all* electrons ($\mu_{Hall}$~60 cm$^2$V$^{-1}$s$^{-1}$) whose concentration is given by $n_{Drude}+n_{peak}$ and includes (*i*) the concentration of the non-equilibrium conduction electrons suffering strong scattering on the quasi-local mode and hence having relatively small mobility ($\mu_{peak}$~15 cm$^2$V$^{-1}$s$^{-1}$) and small mean free path and (*ii*) Drude-type free electrons which are not involved in the formation of the collective excitations and thus characterized by higher mobility values ($\mu_{Drude}$~180 cm$^2$V$^{-1}$s$^{-1}$, Figure 3c).

We suggest that the effect of the discovered non-equilibrium (hot) electrons that make up the majority in LaB$_6$ (up to ≈70%, see Figure 3b) can be considered as the key factor responsible for the extraordinarily low work function of thermoemission in this compound. Taking into account that the highest values of the oscillator strength $f$≈1.64*10$^9$ cm$^{-2}$ (and corresponding electronic concentration), dielectric contribution $\Delta\varepsilon$≈7720 and damping $\gamma$≈1470 cm$^{-1}$ (with corresponding lowest mobility) are observed for La$^{nat}$B$_6$ (Figure 3a) which is characterized by strongest disorder within the boron sublattice, it is natural to conclude that this disorder is among the most important factors determining unique thermoemission characteristics of the LaB$_6$. To unveil microscopic mechanisms responsible for the observed effects low-temperature spectroscopic studies are in progress.

## 4. Conclusions

We have performed broad-band room temperature infrared reflectivity (40 - 35000 cm$^{-1}$), DC-resistivity and Hall-effect experiments on lanthanum hexaborides, which possess record-high thermoemission characteristics. The analysis of the optical conductivity spectra of several isotopically-substituted La($^{10}$B$_x$$^{11}$B$_{1-x}$)$_6$ solid solutions with x=0, 0.189, 0.5, 0.75, 1 allows the conclusion that in addition to the Drude free carrier spectral component, which involves only about 30% of the total number of conduction electrons, there exists an intensive collective mode that is centered at 400-600 cm$^{-1}$ and has a very large dielectric contribution $\Delta\varepsilon$=4000-7200. We argue that the presence in the conduction band of non-equilibrium (hot) electrons involved in the formation of the discovered excitation is at the origin of the extraordinary low electronic work function in LaB$_6$. The finding provides a fresh look at the mechanisms responsible for highest thermoemission characteristics of materials.


**Acknowledgements**

The research was supported by the RSF grant №17-12-01426 and by the Ministry of Education and Science of the Russian Federation (Program 5 top100). Authors acknowledge the Shared Facility Center at P.N. Lebedev Physical Institute of RAS for using their equipment.



**References**

[1] Y. Zhou, C. Fuentes-Hernandez, J. Shim, J. Meyer, A. J. Giordano, Hong Li3, P. Winget, T. Papadopoulos, H. Cheun, J. Kim, M. Fenoll, A. Dindar, W. Haske, E. Najafabadi, T. M. Khan, H. Sojoudi, S. Barlow, S. Graham, J.-L. Brédas, S. R. Marder, A. Kahn, and B. Kippelen, *Science*, **2012**, *336,* 327. DOI: 10.1126/science.1218829





[2] Y. Onuki, A. Umezawa, W. K. Kwok, G. W. Grabtree, M. Nishihara, T. Yamazaki, T. Omi, and T. Komatsubara, *Phys.Rev B.* **1989**, *40*, 11195. DOI: 10.1103/PhysRevB.40.11195

[3] M. Amara, S. E. Luca, R.-M. Galéra, F. Givord, C. Detlefs, and S. Kunii, *Phys. Rev. B,* **2005,** *72*, 064447. DOI: 10.1103/PhysRevB.72.064447

[4] K. Segawa, A. Tomita, K. Iwashita, M. Kasaya, T. Suzuki and S. Kunii, *Journal of Magnetism and Magnetic Materials,* **1992**, *104-107*, 1233. DOI: 10.1016/0304-8853(92)90563-4

[5] M. C. Aronson, J. L. Sarrao, Z. Fisk, M. Whitton, and B. L. Brandt, *Phys Rev. B,* **1999,** *59*, 4720. DOI: 10.1103/PhysRevB.59.4720

[6] J. M.Tarascon, J. Etourneau, P. Dordor, P. Hagenmuller, M. Kasaya, and J. M. D. Coey, *J. Appl. Phys.,* **1980,** *51*, 574. DOI: 10.1063/1.327309

[7] N. Sluchanko, V. Glushkov, S. Demishev, A. Azarevich, M. Anisimov, A. Bogach, V. Voronov, S. Gavrilkin, K. Mitsen, A. Kuznetsov, I. Sannikov, N. Shitsevalova, V. Filipov, M. Kondrin, S. Gabáni, and K. Flachbart, *Phys. Rev. B,* **2017,** *96*, 144501. DOI:10.1103/PhysRevB.96.144501

[8] M. Dzero, K. Sun, V. Galitski, and P. Coleman, *Phys. Rev. Lett.,* **2010,** *104*, 106408. DOI: 10.1103/PhysRevLett.104.106408

[9] T. Mori, *Rare earth borides, carbides and nitrides*, (Ed. D. A. Atwood), John Wiley & Sons Ltd., Chichester, **2012**, p. 263.

[10] K. Iwasa, R. Igarashi, K. Saito, C. Laulh´e, T. Orihara, S. Kunii, K. Kuwahara, H. Nakao, Y. Murakami, F. Iga, M. Sera, S. Tsutsui, H. Uchiyama, and A. Q. R. Baron, *Phys. Rev. B,* **2011,** *84*, 214308. DOI: 10.1103/PhysRevB.84.214308

[11] B. P. Gorshunov, E. S. Zhukova, G. A. Komandin, V. I. Torgashev, A. V. Muratov, Yu. A. Aleshchenko, S. V. Demishev, N. Yu. Shitsevalova, V. B. Filipov, and N. E. Sluchanko, *JETP Letters*, **2018**, *107*, 100. DOI:10.1134/S0021364018020029

[12] N. E. Sluchanko, A. N. Azarevich, A. V. Bogach, N. B. Bolotina, V. V. Glushkov, S. V. Demishev, A. P. Dudka, O. N. Khrykina, V. B. Filipov, and N. Yu. Shitsevalova, *J. Phys.: Condens. Matter,* **2019**, *31*, 065604. DOI: 10.1088/1361-648X/aaf44e





[13] E. S. Zhukova, B. P. Gorshunov, G. A. Komandin, L. N. Alyabyeva, A. V. Muratov, Yu. A. Aleshchenko, M. A. Anisimov, N. Yu. Shitsevalova, S. E. Polovets, V. B. Filipov, and N. E. Sluchanko, arXiv:1811.00104v1.

[14] M. Bakr, R. Kinjo, Y.W. Choi, M. Omer, K. Yoshida, S. Ueda, M. Takasaki, K. Ishida, N. Kimura, T. Sonobe, T. Kii, K. Masuda, H. Ohgaki, and H. Zen, *Phys. Rev. Special Topics - Accelerators and Beams*, **2011**, *14*, 060708, DOI: 10.1103/PhysRevSTAB.14.060708

[15] H. Werheit, V. Filipov, N. Shitsevalova, M. Armbruster and U. Schwarz, *J. Phys.: Condens. Matter*, **2012**, *24*, 385405. DOI: 10.1088/0953-8984/24/38/385405

[16] S.-I. Kimura, T. Nanba, S. Kunii, and T. Kasuya, *J. Phys. Soc. Jpn.,* **1990**, *59*, 3388. DOI: 10.1143/JPSJ.59.3388

[17] H. Okamura, M. Matsunami, T. Inaoka, T. Nanba, S. Kimura, F. Iga, S. Hiura, J. Klijn, and T. Takabatake, *Phys. Rev. B* **62**, R13265 (2000). DOI: 10.1103/PhysRevB.62.R13265

[18] H. Okamura, S. Kimura, H. Shinozaki, T. Nanba, F. Iga, N. Shimizu, and T. Takabatake, *Phys. Rev. B,* **1998**, *58*, R7496. DOI: 10.1103/PhysRevB.58.R7496

[19] M. Dressel, and G. Gruner, *Electrodynamics of Solids,* Cambridge University Press, Cambridge, **2002**.

[20] M. Dressel, N. Kasper, K. Petukhov, B. Gorshunov, G. Gruner, M. Huth, and H. Adrian, *Phys. Rev. Lett.*, **2002**, *88*, 186404. DOI: 10.1103/PhysRevLett.88.186404; D. N. Basov, R.D. Averitt, D. van der Marel, M. Dressel, and K. Haule, Rev. Mod. Phys. **83**, 471 (2011). DOI: 10.1103/RevModPhys.83.471.

[21] N. B. Bolotina, A. P. Dudka, O. N. Khrykina, V. V. Glushkov, A. N. Azarevich, V. N. Krasnorussky, S. Gabani, N. Yu. Shitsevalova, A. V. Dukhnenko, V. B. Filipov, N. E. Sluchanko, arXiv:1811.06441.

[22] M. A. Anisimov, V. V. Glushkov, A. V. Bogach, S. V. Demishev, N. A. Samarin, S. Yu. Gavrilkin, K. V. Mitsen, N. Yu. Shitsevalova, A. V. Levchenko, V. B. Filipov, S. Gab´ani, K. Flachbart, and N. E. Sluchanko, *JETP,* **2013**, *116*, 760. DOI: 10.1134/S1063776113050014





[23] Y. Ishizawa, T. Tanaka, E. Bannai, and S. Kawai, *J. Phys. Soc. Jpn.,* **1977**, *42*, 112. DOI: 10.1143/JPSJ.42.112




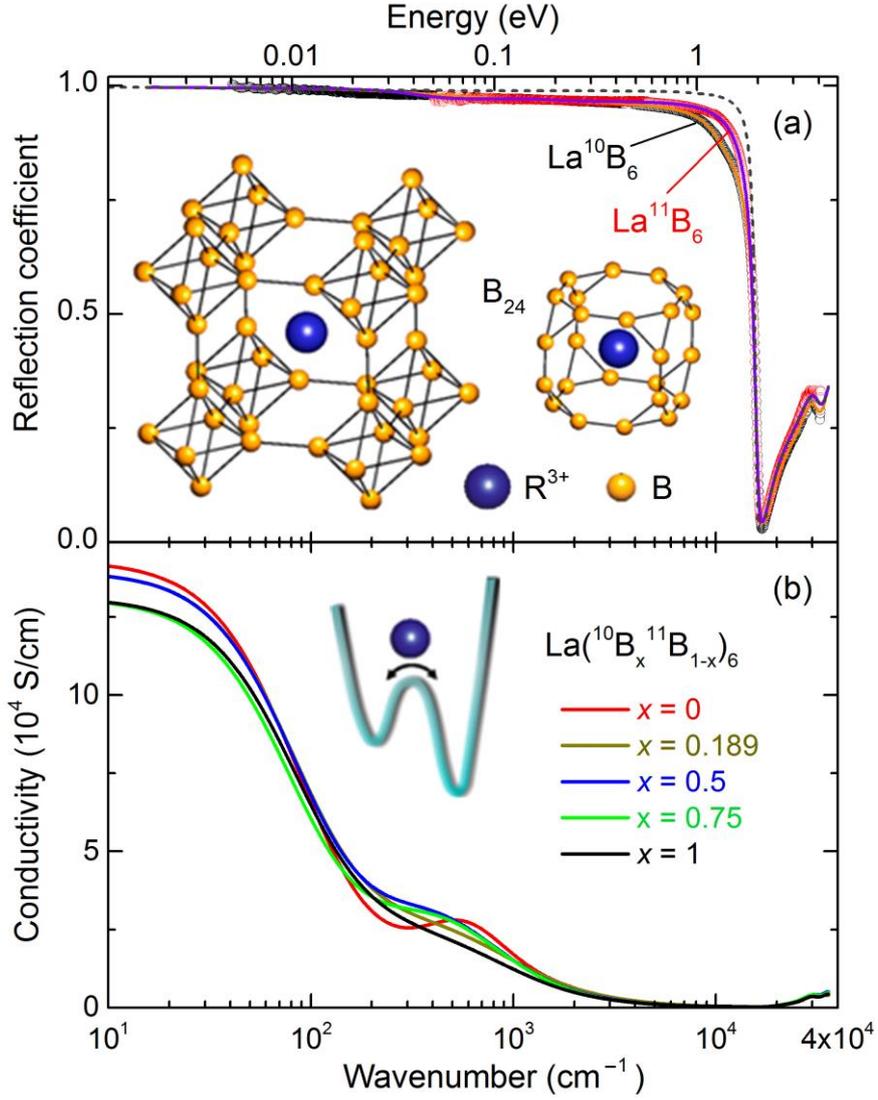

**Figure 1.** (Color online). (a) Reflectivity spectra of La$^{10}$B$_6$ and La$^{11}$B$_6$ single crystals (dots). Solid lines show the results of least-square fitting of the spectra with the Drude term, Equation 1, and two Lorentzian terms, Equation 2, as described in the text. The dashed line shows the best fit of the reflectivity spectrum of La$^{11}$B$_6$ obtained by the Drude term, Equation 1, ($\sigma_{DC}^{Drude}$=143110 Ohm$^{-1}$cm$^{-1}$ and $\gamma^{Drude}$=260 cm$^{-1}$). The obtained spectra of the real part of conductivity of all measured La($^{10}$B$_x$$^{11}$B$_{1-x}$)$_6$ crystals are presented in panel (b). The insets in panels present the crystal structure of RB$_6$ and Fedorov B$_{24}$ polyhedra centered by R$^{3+}$ ion and vibrations of La- ion in the double-well potential (schematically). The data are recorded at ambient temperature $T$=300 K.

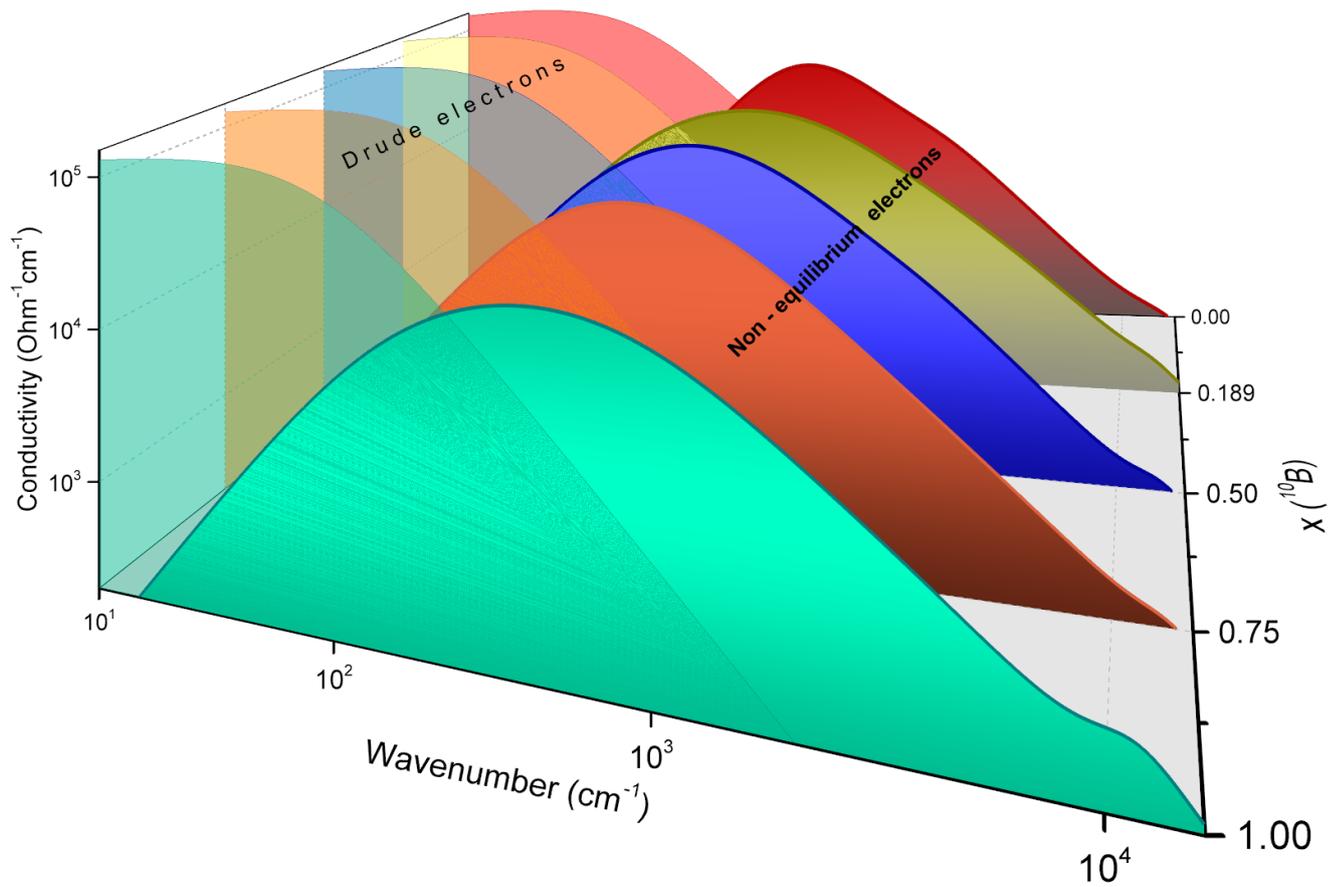

**Figure 2.** (Color online). Contributions observed in the infrared conductivity spectra of La($^{10}$B$_x$$^{11}$B$_{1-x}$)$_6$ single crystals from Drude-type electrons in the conduction band (Drude electrons) and collective peak (non-equilibrium electrons). The temperature is $T$=300 K. Note the logarithmic scales in frequency and conductivity.



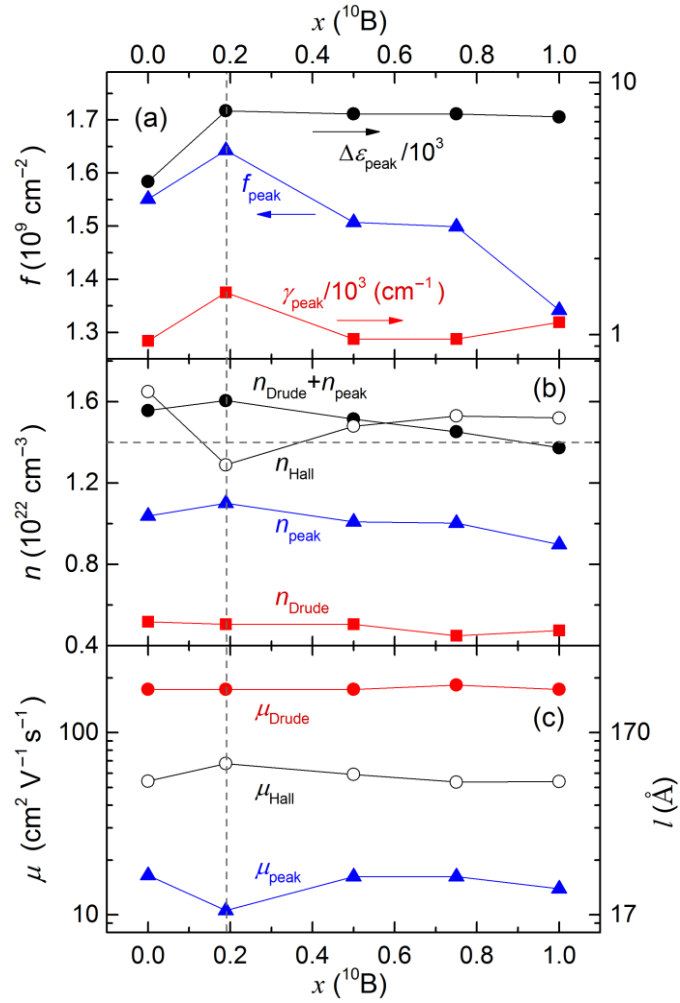

**Figure 3.** (a) Dependences on the relative contents of $^{10}$B and $^{11}$B isotopes in La($^{10}$B$_x$$^{11}$B$_{1-x}$)$_6$ single crystals of the parameters of infrared collective peak: dielectric contribution $\Delta\varepsilon_{peak}$, oscillator strength $f_{peak}$ and damping $\gamma_{peak}$. Panel (b) shows the same dependences for free charge carriers concentration $n_{Drude}$, concentration $n_{peak}$ of charge carriers responsible for the formation of collective absorption peak, combined concentration $n_{Drude} + n_{peak}$ and the concentration $n_{Hall}$ of carriers obtained from Hall measurements. Horizontal dashed line corresponds to the number of La$^{3+}$ ions per unit volume in LaB$_6$ $n$(La)~1.4*10$^{22}$ cm$^{-3}$. (c) Dependence on $x(^{10}B)$ of charge carriers' mobility and mean-free path obtained from Hall measurements and from optical experiments ($\mu_{Drude}$, $\mu_{peak}$, respectively). Note logarithmic scale on the left axis. Vertical dashed line denotes points corresponding to the natural compound $^{nat}$LaB$_6$. Temperature $T$=300 K.